\documentclass[a4paper]{jpconf}
\usepackage{graphicx}
\def\sqrtsNN{\mbox{$\sqrt{s_\mathrm{NN}}$}}

\newcommand{ \be }{\begin{equation}}    
\newcommand{ \ee }{\end{equation}}    
\newcommand{ \bea }{\begin{eqnarray}}    
\newcommand{ \eea }{\end{eqnarray}}

\begin{document}
\title{Beam Energy Dependence of Clan Multiplicity at RHIC}

\author{Aihong Tang (for the STAR Collaboration)}

\address{Physics Department, Brookhaven National Laboratory, PO BOX 5000, Upton, NY 11973-5000}

\ead{aihong@bnl.gov}

\begin{abstract}
In this paper, STAR's measurement of clan multiplicity is presented for AuAu collisions at \sqrtsNN = 7.7, 11.5, 19.6, 27, 39, 62.4 and 200 GeV, for a variety of centrality classes. The mean number of particles per clan is found to decrease with decreasing centrality. Within the same centrality class, the mean number of particles per clan exhibits a reduction between 19.6 GeV and 62 GeV, with the minimum around 27 GeV. The structure is visible for most centralities, and most prominent for central collisions. 
\end{abstract}

\section{Introduction}

Particle distributions in high energy collisions are frequently found to be of Negative Binomial Distribution (NBD) form for both total multiplicities and multiplicities in finite pseudorapidity intervals. One satisfactory explanation of this observation is that particle are produced via cascade process, and such interpretation leads to the  development of clan concept of clustering in particle production~\cite{Giovanni86}. 

Let's recall the negative binomial probability density function,
\be   
f(n)= {k+n-1 \choose n}(1-p)^n p^k,
\label{eq:NegBinom}
\ee   
where $p \in [0,1]$, and the real number $k>0$. 
Its mean and variance are given by $\mu=\frac{(1-p)k}{p}$ and $\sigma^2=\frac{(1-p)k}{p^2}$, respectively. In order to interpret the wide occurrence of NBD, a parameterization alternative to the standard NBD parameterization has been proposed~\cite{Giovanni86}. It consists of the average number of groups of particles of common ancestor, also called the average number of clans,
\be   
\bar{N}=k \mathrm{ln} (1+\frac{\mu}{k}),
\label{eq:meanN}
\ee   
and the average number of particles per clan,
\be   
\bar{n}_c=\frac{\mu}{\bar{N}}.
\ee   
With this set of parameterization, particle production can be regarded as an independent production of clans followed by the decay of clans into particles according to a logarithmic multiplicity distribution. This can be seen by investigating the generating functions of relevant distributions. 
The generating function for the logarithmic distribution is,
\be   
G_{\mathrm{ln}}(x) = \frac{\mathrm{ln}(1-zx)}{\mathrm{ln}(1-z)},
\ee   
where $z$ is set to be $z=\frac{\mu}{\mu +k}$. Because clans are independently produced, they can be described by Poisson distribution, for which the generating function is
\be   
G_{\mathrm{Poisson}}(x) = e^{\bar{N}(x-1)}.
\ee   
The final distribution is a compound distribution of logarithm and Poisson, for which the generating function is given by
\bea   
G_{\mathrm{compound}} &=& G_{\mathrm{Poisson}}(G_{\mathrm{ln}}(x)) \nonumber \\
&=& e^{\bar{N}(G_{\mathrm{ln}(x)}-1)} \nonumber \\
&=&  \left( \frac{1-z}{1-zx} \right )^k,
\eea 
by which we have recovered the generating function of NBD.

Note that with the introduction of the clan concept, the NBD, if arranged properly, yields a similarity to the grand-canonical partition function. Such analogy led to the development of clan thermodynamics~\cite{clanThermal02, clanThermal08}. Here, adopting the concept of clan does not necessarily imply the endorsement of clan thermodynamics.  The clan parameters have been used to identify abnormalities due to phase transition~\cite{UA5, EMC, NA22, E802, NA35, PHENIX}. For a similar purpose, this study is to examine the averaged charged multiplicity per clan as a function of collision energy.

\section{Data Sets and Cuts}
The data set consist of minimum bias events of Au+Au collisions at \sqrtsNN = 7.7, 11.5, 19.6, 27, 39, 62.4 and 200 GeV, taken by the STAR experiment. Total events used in this analysis are, from the lowest energy to the highest one, 5, 7, 40, 71, 56, 73, and 25 million, respectively. Events from energies lower than 62.4 GeV were taken as part of the Beam Energy Scan program~\cite{BESWhitePaper} at RHIC, and events from 62.4 and 200 GeV were taken in year 2010. All events are required to have at least one Time of Flight~\cite{TOF} hit matching to a track,  a vertex within 30 centimeters to the center of STAR's Time Projection Chamber~\cite{TPC} in z direction, and within 2 centimeters to the beam in transverse direction. To avoid complications arising from having more than one collisions in the time window of detector readout, which usually happens when the luminosity is high, events from 62.4 GeV and 200 GeV are selected from those taken with low luminosity beam (Beam Beam Counter coincidence rate less than 25k). Tracks are required to have a minimum of 15 hits, and the ratio of number of hits to number of maximum possible hits to be greater than 0.52. The distance of the closest approach (DCA) of a track to the primary vertex is required to be less than 1 centimeter. To avoid auto correlation, particles used in the study of multiplicity distribution and those used in the centrality definition are taken from different pseudorapidity ranges, which is $|\eta|<0.5$ for the former and $|\eta|>0.5$ for the latter. In this paper the multiplicity in the latter is also called ``reference-multiplicity-2" to distinguish itself from the usual reference multiplicity defined with particles in $|\eta|<0.5$. In the picture of particles being produced by clans, one is interested in produced particles only. Thus protons are removed in this study, because they are contaminated by transported ones which can introduce trivial energy dependence, in particular at low energies where transported protons become dominant. The removal of protons is achieved by requiring tracks' energy loss per unit length ($dE/dx$) to be larger than a threshold set to 2 standard deviations below the theoretically expected value for protons at the same rigidity.

\section{Analysis Procedure and Results}
\begin{figure}[h]
\begin{center}
\includegraphics[width=20pc]{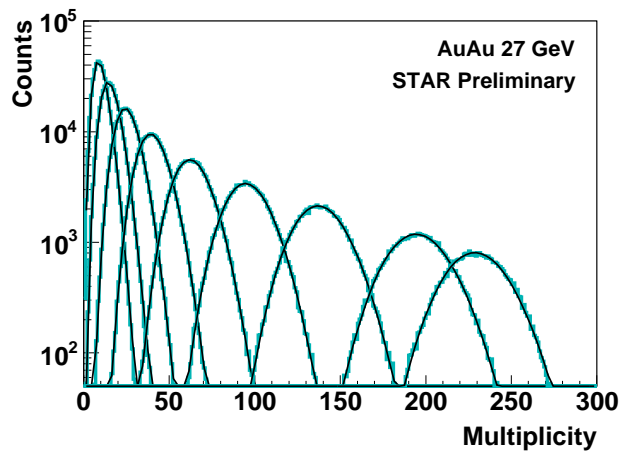}
\caption{\label{fig:NBD_data}From left to right, the multiplicity distribution in $|\eta|<0.5$ for reference-multiplicity-2 = 11, 18, 30, 48, 76, 116, 169, 242 and 289, respectively. Curves are fits to NBD. All histograms are for Au+Au collisions at  \sqrtsNN = 27 GeV. }
\end{center}
\end{figure}

For events with a given reference-multiplicity-2, the corresponding multiplicity distribution in $|\eta|<0.5$ can be well described by NBD. This can be seen in Figure~\ref{fig:NBD_data}, in which the multiplicity distribution in $|\eta|<0.5$ is plotted for a selected set of reference-multiplicity-2 numbers. The curves are fits to the NBD and they describe the data very well. 

\begin{figure}[h]
\begin{center}
\includegraphics[width=20pc]{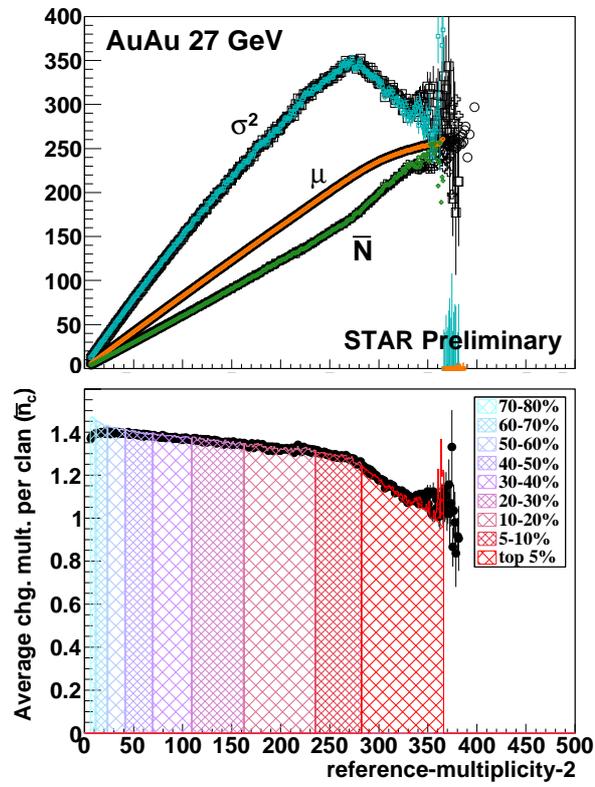}
\caption{\label{fig:fitPar_refMult2} $\sigma^2$, $\mu$, $\bar{N}$ (top panel) and $\bar{n}_c$ (bottom panel) as a function of reference-multiplicity-2 for Au+Au collisions at  \sqrtsNN = 27 GeV. See text for details.}
\end{center}
\end{figure}

The fit is applied to multiplicity distributions with the finest bin of reference-multiplicity-2, and the corresponding $\sigma^2$ and $\mu$ can be obtained. With both $\sigma^2$ and $\mu$ known, $\bar{N}$ and $\bar{n}_c$ can be calculated with the transformation mentioned above. In the top panel of Figure~\ref{fig:fitPar_refMult2},  $\sigma^2$, $\mu$ and $\bar{N}$ are plotted as a function of reference-multiplicity-2, and in the bottom panel, the average multiplicity for charged particles per clan ($\bar{n}_c$) is shown. Note that $\bar{N}$ and $\bar{n}_c$ can be calculated with either directly calculated $\sigma^2$ and $\mu$, shown by black symbols, or that from fitting the NBD, shown by non-black symbols. The two sets of results agree with each other well and in the follow-up analysis only the set based on fitting NBD is used, because fitting is less sensitive to contaminations from rare events with abnormally large or small multiplicity. In general $\sigma^2$ is found to be greater than $\mu$ which is a feature of NBD. With decreasing centrality (increasing reference-multiplicity-2), $\bar{N}$ keep increasing while $\bar{n}_c$ shows a moderate decrease towards unit, which means that particle production is approaching the limit of independent production (Poisson) in central collisions.  

The obtained $\bar{N}$ and $\bar{n}_c$ are then averaged over reference-multiplicity-2 for each centrality bin, of which the boundaries are shown by mesh area in the bottom panel of Figure~\ref{fig:fitPar_refMult2}. Note by performing the analysis with the finest reference-multiplicity-2 bin, one avoids the additional fluctuation from a wide multiplicity bin which can significantly bias the variance. 

\begin{figure}[h]
\begin{center}
\includegraphics[width=23pc]{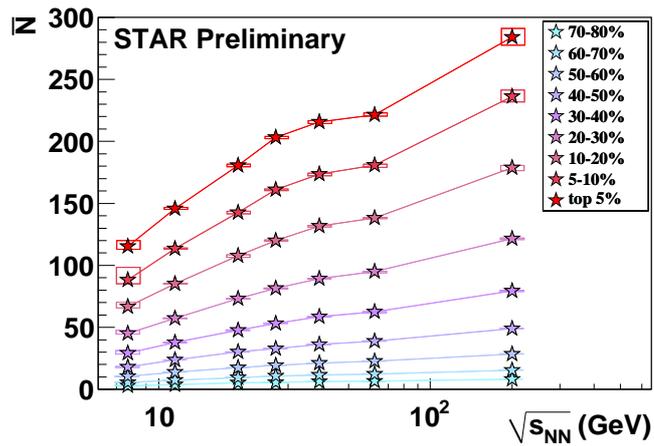}
\caption{\label{fig:meanNClan} Average total number of clans per event $\bar{N}$ as a function of collision energy, for various centrality classes. Boxes represents systematic error resulted from variations in STAR's day-to-day operation during data-taking.}
\end{center}
\end{figure}

\begin{figure}[h]
\begin{center}
\includegraphics[width=23pc]{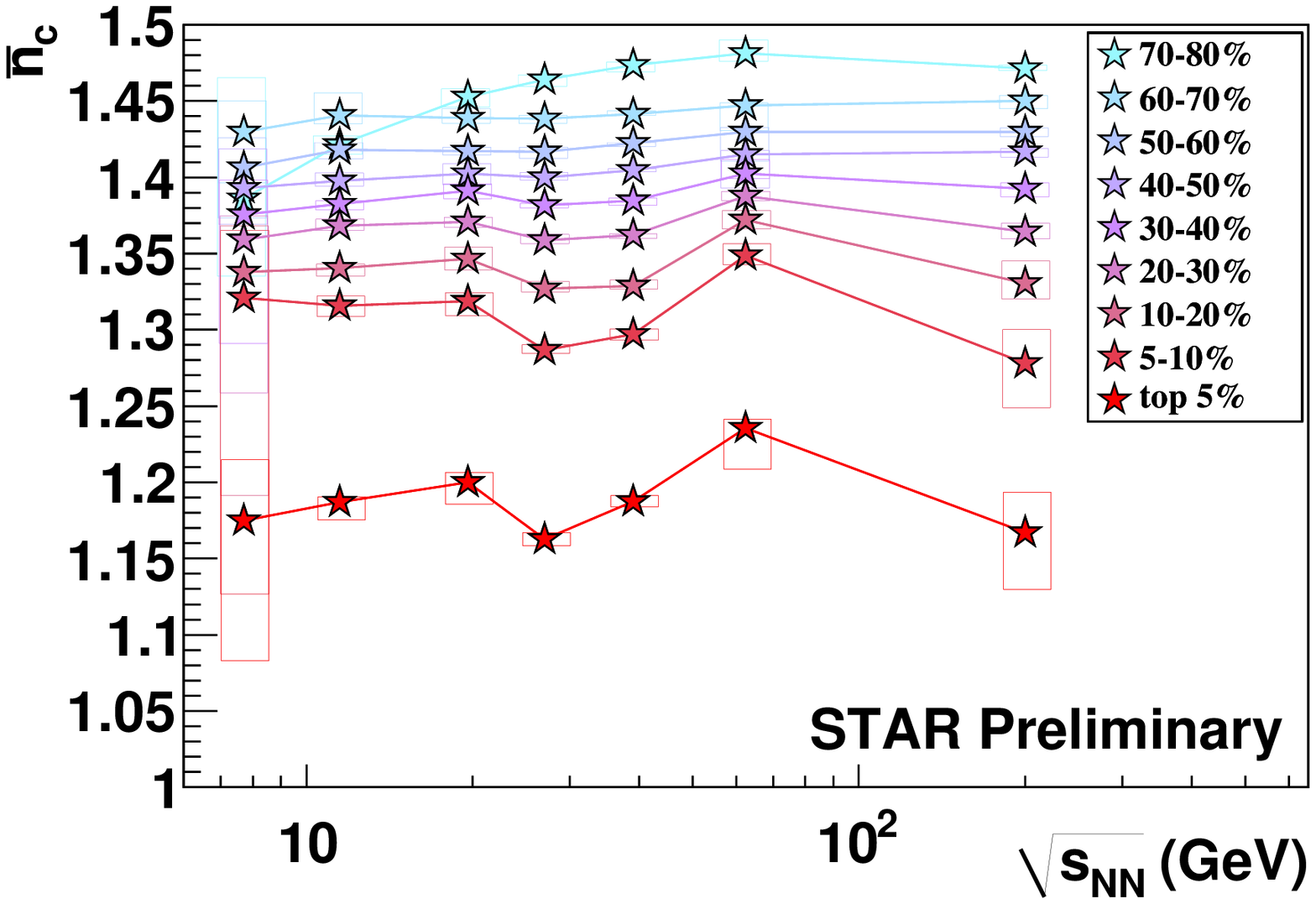}
\caption{\label{fig:meanClanM} Average charged particles per clan $\bar{n}_c$ as a function of collision energy, for various centrality classes. Boxes represents systematic error resulted from variations in STAR's day-to-day operation during data-taking.}
\end{center}
\end{figure}

The same procedure is applied on data from all energies and the final $\bar{N}$ and $\bar{n}_c$ as a function of collision energy is presented in Figure~\ref{fig:meanNClan} and Figure~\ref{fig:meanClanM}, respectively. Not surprisingly the average total number of clans per event is found to increase either with decreasing centrality, or with increasing energy, indicating that more clans are produced when the system becomes larger and/or when there is more energy available for particle production. The average charged particles per clan shows little dependence on energy for peripheral collisions, which is expected if peripheral collisions are regarded as simple superpositions of independent binary collisions, thus they should show the same clustering property. From peripheral to central collisions, $\bar{n}_c$ exhibits an energy dependence, with a minimum at 27 GeV and an enhancement at 62 GeV. Such dependence is seen in non-peripheral collisions and is strongest in most central collisions. If there is a phase transition, one would expect a disturbance in particle production which reveals itself through the change of clustering characteristics. The feature observed is a motivation for further studies in this direction.  It is worth to point out that the dependence becomes less significant when the DCA cut is released to 2 cm or 3 cm, or when protons, contaminated by transported ones, are included. Both observations (not shown) indicate that the feature is prominent in primarily produced particles, for which the clan concept applies. The apparent enhancement at 62 GeV may be affected by worse beam condition at that particular energy and needs a further investigation.

Due to the change in operation condition, for example, temperature and air pressure of TPC, the efficiency is not a constant. Nevertheless, one can correct for efficiency and recover the efficiency-corrected $\bar{N}$ and $\bar{n}_c$ by applying the following two identities~\cite{TangWang}:
\bea   
\mu &=& \frac{\mu'} {\epsilon} \nonumber \\
p &=& \frac{p'\epsilon}{1-p'(1-\epsilon)} ,
\eea 
where $\mu'$ and $p'$ are measured observables, and $\mu$ and $p$ are efficiency-corrected ones. Here $\epsilon$ is the {\it detecting efficiency} in general meaning, which includes both the finite detector efficiency effect and the finite detector acceptance effect, $\epsilon$ = efficiency $\times$ acceptance. With that, the efficiency corrected $\bar{N}$ and $\bar{n}_c$ can be calculated,
\bea   
\bar{N} &=& -k \mathrm{ln}p = -k \mathrm{ln} \big[ \frac{p'\epsilon}{1-p'(1-\epsilon)} \big] \nonumber \\
\bar{n}_c &=& \frac{\mu}{\bar{N}} = \frac{\mu'/\epsilon}{-k \mathrm{ln} \big[ \frac{p'\epsilon}{1-p'(1-\epsilon)}\big]}.
\eea 
To study if the efficiency variation can cause such a feature, without loosing generality, assuming a 30\% efficiency and 5\% variation in year-to-year operation, the resulted variation in $\bar{n}_c$ is too small (comparable to symbol size in Figure~\ref{fig:meanClanM}) to explain the variation seen in data. 
 
\section{Summary}
In this paper, STAR's measurement of average total number of clans per event, $\bar{N}$, and average charged particles per clan, $\bar{n}_c$, are studied for AuAu collisions at \sqrtsNN = 7.7, 11.5, 19.6, 27, 39, 62.4 and 200 GeV, for a variety of centrality classes. $\bar{N}$ is found to increase with decreasing centrality and/or increasing energy, and $\bar{n}_c$ is found to decrease with decreasing centrality. Within the same centrality class, $\bar{n}_c$ exhibits a reduction between 19.6 GeV and 62 GeV, with the minimum around 27 GeV. The structure is not seen in peripheral collisions (50\%-70\%) but is visible in middle central collisions and, becomes prominent in central collisions (top 5\%). The variation of mean number of particles per clan with energy cannot be explained by variation in detecting efficiency. Whether the observed structure is connected to a phase transition needs further investigations.

\end{document}